# Structure-coupled joint inversion of geophysical and hydrological data

Tobias Lochbühler*[1], Joseph Doetsch[2], Ralf Brauchler[3], Niklas Linde[1]

[1]Faculty of Geosciences and Environment, University of Lausanne, UNIL Mouline – Géopolis, 1015 Lausanne, Switzerland

[2]Earth Sciences Division, Lawrence Berkeley National Laboratory, One Cyclotron Road, Berkeley, CA 94720, USA

[3]Department of Earth Sciences, Swiss Federal Institute of Technology, Sonneggstrasse 5, 8092 Zurich, Switzerland

*Corresponding author: Tobias.Lochbuehler@unil.ch

# ABSTRACT

In groundwater hydrology, geophysical imaging holds considerable promise for improving parameter estimation, due to the generally high resolution and spatial coverage of geophysical data. However, inversion of geophysical data alone cannot unveil the distribution of hydraulic conductivity. Jointly inverting geophysical and hydrological data allows benefitting from the advantages of geophysical imaging and, at the same time, recover the hydrological parameters of interest. We introduce a first-time application of a coupling strategy between geophysical and hydrological models that is based on structural similarity constraints. Model combinations, for which the spatial gradients of the inferred parameter fields are not aligned in parallel, are penalized in the inversion. This structural coupling does not require introducing a potentially weak, unknown and non-stationary petrophysical relation to link the models. The method is first tested on synthetic data sets and then applied to two combinations of geophysical/hydrological data sets from a saturated gravel aquifer in northern Switzerland. Crosshole ground-penetrating radar (GPR) travel times are jointly inverted with hydraulic tomography data, as well as with tracer mean arrival times, to retrieve the 2-D distribution of both GPR velocities and hydraulic conductivities. In the synthetic case, it is shown that incorporating the GPR data through a joint inversion framework can improve the resolution and localization properties of the estimated hydraulic conductivity field. For the field study, recovered hydraulic conductivities are in general agreement with flowmeter data.

# INTRODUCTION

Geophysical imaging has become a popular component of parameter estimation procedures in groundwater hydrology (e.g., Rubin et al., 1992; Hubbard et al., 2001; Singha and Gorelick, 2005; Rubin and Hubbard, 2005; Dafflon and Barrash, 2012). Geophysical experiments may provide large data sets of high spatial density and inversion of these data leads to images that capture the spatial distribution of the underlying geophysical property at a relatively high resolution. However, geophysical imaging primarily provides information about properties that the measured geophysical data are directly sensitive to, for example, acoustic velocity in seismic tomography, radar velocity in ground-penetrating radar (GPR) tomography, or electrical resistivity in electrical resistance tomography (ERT). The resulting geophysical models often allow for structural interpretation, such as interface localization or detection of faults, paleo-channels or aquifer boundaries (e.g., Hyndman and Tronicke, 2005). But, reliable modeling of groundwater flow and solute transport requires detailed quantitative information about the distribution of hydraulic conductivity, which geophysics alone cannot provide (e.g., Yeh et al., 2008). Classical geophysical imaging hence provides high-resolution images of structures that may be essential for predictive numerical groundwater modeling, but generally fails to produce the numerical values of the hydrological parameters of interest.

In the last two decades, various approaches have been developed that aim to retrieve detailed hydraulic models from geophysical data or from combinations of geophysical and hydrological data. Most early methods were based on first inverting the geophysical data to estimate the spatial distribution of a geophysical property, which in turn was translated into a hydrological parameter field using petrophysical relations and sometimes site-specific hydrological data (e.g., Hubbard and Rubin, 2000). In a case study, Hubbard et al. (2001) found a correlation coefficient between the inferred radar velocity and the log hydraulic conductivity to be close to 0.7. Such a strong correlation might justify a site-specific petrophysical model to link the geophysical model estimates and the hydrological properties. Generally speaking, this approach is problematic since the necessary petrophysical link is often poorly known, non-stationary in space and time and scale-dependent (Day-Lewis et al., 2005; Pride, 2005). More recently, the problem of site-specific petrophysics was tackled by including the petrophysical parameters in the inversion and updating them together with the hydrological model until the geophysical data are matched (or vice versa) (e.g., Kowalsky et al., 2005; Hinnell et al., 2010; Linde et al., 2006b). Another approach is to infer the geometry of hydrogeological zones of approximately uniform properties (so-called hydrofacies, Anderson (1989)) based on the geophysical data and to estimate the hydrological properties of the individual zones by inverting hydrological data (e.g., Hyndman et al., 1994; Hyndman and Gorelick, 1996). Similarly, Eppstein and Dougherty (1998) used zones of high and low GPR velocities as qualitative indicators of low and high moisture content.

Any explicit petrophysical coupling between the hydrological and the geophysical models within a joint inversion process carries the risk of propagating errors produced by an overly simplified or incorrect petrophysical relation to the resulting hydrological model (even if the parameters of the petrophysical relations are updated, the parametric form might be biased or too simplistic). Consequently, any hydrological predictions that are based on this model are prone to be biased. This limitation can be overcome by using structure-coupled joint inversions that do not assume an explicit petrophysical relation between the properties inverted for. The different models are instead coupled by assuming that the spatial distributions of the model parameters have similar patterns within the model domain (Haber and Oldenburg, 1997; Gallardo and Meju, 2004). Cross-gradient joint inversions are nowadays used rather widely in geophysics to combine data sets acquired with different geophysical techniques (e.g., Gallardo and Meju, 2003; 2004; 2011; Tryggvason and Linde, 2006; Linde et al., 2006a; 2008; Doetsch et al., 2010; Moorkamp et al., 2011; Gallardo et al., 2012). In this approach, structural differences between the models are quantified by the cross product of the spatial gradients of the model parameter fields. Structural similarity is then enforced by penalizing model combinations for which this cross gradient function is non-zero at any location in the model domain. The assumption of structural similarity between the models is based on the generally strong conditioning of geophysical property variations by geological structures. Different geophysical properties often map the same lithological units or change across the same boundaries. Since hydrological properties generally change with the lithofacies, a strong structural link between the geophysical and hydrological parameter fields is expected. Bayer et al. (2011), for example, found the lithofacies distribution imaged with surface GPR and the hydrofacies distribution in an alluvial aquifer to be closely related (see also Kowalsky et al., 2001). Similarly, Dogan et al. (2011) combined direct push measurements of hydraulic conductivity with 3-D GPR imaging and observed that the hydraulic conductivity variations are much smaller within units detected by the GPR experiments than between them. The assumption of structural similarity between models is violated for scenarios where the applied methods are not sensitive to the same structural information, or where one method is primarily sensitive to lithological structures, and the other to the dynamic system state, such as saturation or salinity (see discussion in Linde et al. (2006a)).

For the first time, we here adapt the structure-coupled joint inversion approach to invert combinations of geophysical and hydrological data sets to better resolve the subsurface distribution of hydraulic conductivity. The main potential advantages of this approach are that it enables us (1) to use the high-resolution geophysical data to improve the spatial resolution of the hydraulic model and (2) to avoid introducing any petrophysical relation (other than structural similarity) that relates the geophysical and the hydraulic models. Two different examples are considered: (a) joint inversion of travel times and amplitude attenuation data derived from multi-level crosshole slug interference tests (hydraulic tomography data) together with crosshole GPR travel times; and (b) joint inversion of tracer mean arrival times derived from bank filtration in a river-groundwater system and crosshole GPR travel times.

In both cases, the goal is to retrieve 2-D models of hydraulic conductivity and radar velocity in a saturated porous aquifer. Our inversion approach builds on the work by Linde et al. (2006a) and Doetsch et al. (2010). They developed an algorithm for structure-coupled joint inversion of multiple geophysical data sets, which was strongly influenced by the pioneering work of Gallardo and Meju (2003; 2004). We apply our method to a synthetic example and to field data from a gravel aquifer in the Thur valley in northern Switzerland. The main objectives of this study are twofold: (1) to assess the improvement in hydrological parameter estimation obtained by incorporating geophysical data compared to solely inverting hydrological data and (2) to test the performance of structure-coupled joint inversion for combinations of geophysical and hydrological data in a typical field setting.

The structure of this paper is as follows: In the first two parts, we describe the hydrological and geophysical methods used and demonstrate how they are combined in the joint inversion framework. We then present results of a synthetic test example and the detailed field study, followed by a discussion of our findings and a summary of the benefits and limitations of the method.

# HYDROLOGICAL AND GEOPHYSICAL METHODS

## Multi-level crosshole slug interference tests

Usually, slug tests are conducted to get an estimate of the local hydraulic conductivity around a borehole. If performed in crosshole mode, slug tests can provide information about the hydraulic conductivity and specific storage between the boreholes. For this type of experiments, double-packer systems are used to hydraulically isolate sections of the test and observation well. Hydraulic crosshole slug interference tests are then performed between these isolated sections. Varying the positions of the double-packer systems in the test and observation well allows sensing different parts of the interwell region (Butler et al., 1999; Yeh and Liu, 2000). Following Brauchler et al. (2011), we reconstruct the hydraulic diffusivity ($D$ [m$^2$/s]) and the specific storage ($S_s$ [1/m]) distribution between test and observation wells by hydraulic travel time and hydraulic attenuation tomography. The hydraulic conductivity ($K$ [m/s]) distribution is then calculated as

$$K_i = D_i S_{s,i}, \qquad (1)$$

where $i$ depicts the index of the grid cell.

The hydraulic tomographic forward model is based on the transformation of the diffusivity equation into the eikonal equation using an asymptotic approach (Virieux et al., 1994; Vasco et al., 2000). The applicability of this ray approximation of the pressure pulse has been proven for synthetic examples and field studies by various authors (Brauchler et al., 2003; 2011; Vasco et al., 2000; Kulkarni et al., 2001; He et al., 2006; Hu et al., 2011). The eikonal equation can be solved efficiently with ray-tracing techniques or particle tracking methods. In this study, the eikonal equation is solved with the finite difference algorithm of Podvin and Lecomte (1991) within the nonlinear travel time tomography algorithm *ps_tomo* (Tryggvason et al., 2002) to provide hydraulic travel times and attenuations.

In the following, we give a short description of the hydraulic travel time and hydraulic attenuation inversion following Vasco et al. (2000) and Brauchler et al. (2003; 2011). In hydraulic travel time tomography the following line integral, introduced by Vasco et al. (2000), is solved

$$\sqrt{t_{\text{peak}}(\mathbf{x}_2)} = \frac{1}{\sqrt{6}} \int_{\mathbf{x}_1}^{\mathbf{x}_2} \frac{ds}{\sqrt{D(s)}}, \qquad (2)$$

where $D(s)$ is the hydraulic diffusivity as a function of arc length $s$ [m] along the propagation path going from $\mathbf{x}_1$ to $\mathbf{x}_2$ and $t_{peak}$ [s] is defined as the time of the peak magnitude in the recorded transient pressure curve in the observation interval.

Similar to equation 2, Brauchler et al. (2011) defined a line integral, which relates the attenuation of a transient hydraulic pressure signal to the specific storage. The attenuation is defined as the ratio of the hydraulic head at the observation interval $h(\mathbf{x}_2)$ [m] and the initial displacement $H_0$ [m] measured in the test interval. Assuming a Dirac pulse source signal, the amplitude decay is described as

$$\left(\frac{h(\mathbf{x}_2)}{H_0}\right)^{-\frac{1}{3}} = B^{-\frac{1}{3}} \int_{\mathbf{x}_1}^{\mathbf{x}_2} \left(\frac{1}{S_s(s)}\right)^{-\frac{1}{3}} ds, \qquad (3)$$

where $S_s(s)$ is the specific storage depending on the arc length $s$ and $B$ is a test specific parameter defined as

$$B = \frac{\pi r_c^2}{\sqrt{\left(\frac{2\pi}{3}\right)^3}} \exp\left[-\frac{3}{2}\right], \qquad (4)$$

where $r_c$ is the radius of the well casing.

**Temporal moments of tracer breakthrough curves**

The advantages of inverting temporal moments of tracer breakthrough curves for estimating hydraulic conductivity have been exposed by various authors (e.g., Harvey and Gorelick, 1995; Cirpka and Kitanidis, 2000a). Unlike modeling of the concentration evolution over time (e.g., Hyndman et al., 1994), temporal moments are described by steady-state equations. This makes the forward modeling process much faster, since it does not require transient transport modeling and has the additional advantage that the moment representation reduces the number of data in the inversion process. We refer to Leube et al. (2012) for a demonstration of temporal moments' compressive capacity to represent information in breakthrough curves. Recently, Pollock and Cirpka (2012) applied a moment-based approach in geophysics by inverting time-lapse ERT data of a tracer experiment conducted in a sandbox.

The $k$-th temporal moment $\mu_k$ of a breakthrough curve $c(\mathbf{x},t)$ acquired at position $\mathbf{x}$ is defined as

$$\mu_k(\mathbf{x}) = \int_0^\infty t^k c(\mathbf{x},t) dt . \tag{5}$$

From this equation, moment-generating equations are derived by multiplying the advection-dispersion equation for $c$ with $t^k$ and integrating over time, applying integration by parts (Harvey and Gorelick, 1995; Cirpka and Kitanidis, 2000b). The resulting moment-generating equations are partial differential equations of the form (e.g., Nowak and Cirpka, 2006)

$$\nabla \cdot (\mathbf{v}\mu_k - \mathbf{D}\nabla \mu_k) = k\mu_{k-1}, \tag{6}$$

subject to the boundary conditions:

$$(\mathbf{v}\mu_k - \mathbf{D}\nabla \mu_k) \cdot \mathbf{n} = \mathbf{v} \cdot \mathbf{n}\hat{\mu}_k \quad \text{on the inflow boundary and} \tag{7}$$

$$(\mathbf{D}\nabla \mu_k) \cdot \mathbf{n} = 0 \quad \text{on the outflow boundary.} \tag{8}$$

Here, $\mathbf{v}$ is the velocity vector, $\mathbf{n}$ is a unit vector normal on the boundary and $\hat{\mu}$ is a specified moment. $\mathbf{D}$ is the dispersion tensor given as (Scheidegger, 1961)

$$D_{ij} = \frac{v_i v_j}{\|\mathbf{v}\|}\left[(\alpha_l - \alpha_t) + \delta_{ij}(\alpha_t \|\mathbf{v}\| + D_m)\right], \tag{9}$$

where $v_i$ is the $i$-th component of the velocity vector, $\alpha_l$ and $\alpha_t$ [m] are the longitudinal and transversal dispersivities, respectively. $D_m$ is the molecular diffusion coefficient [m$^2$/s] and $\delta_{ij}$ is the Kronecker symbol, which is 1 for $i = j$ and 0 for $i \neq j$. The dependency on the hydraulic conductivity $K$ and the porosity $\varphi$ within each grid cell arises from Darcy's law for the seepage velocity

$$\mathbf{v} = -\frac{1}{\varphi} K \nabla h. \tag{10}$$

A detailed description of the physical meaning of the different temporal moments is given by Harvey and Gorelick (1995) and Cirpka and Kitanidis (2000a). In our approach, we only use the first moment normalized with the zeroth moment. For a Dirac-type tracer injection, the normalized first moment $\frac{\mu_1}{\mu_0}$ [s] of a breakthrough curve is the mean arrival time of a tracer (Cirpka and Kitanidis, 2000b). We will omit the normalization in the following since $\mu_0 = 1$ for a Dirac-type injection. Higher order moments carry spatial information about the dispersion characteristics of the system. Here, only first normalized moments are considered as data since they are strongly sensitive to hydraulic conductivity, which is what we are primarily interested in, but also because the estimates of higher moments are less precise. For the forward calculation of the moments, we solve equation 6 on a regular finite element mesh, using the code of Nowak (2005).

Apart from the velocity field, which is controlled by the hydraulic conductivity model and the pressure gradient, the characteristics of the tracer breakthrough are influenced by dispersion phenomena, which are taken into account in our models by considering the molecular diffusion coefficient $D_m$ and the dispersivities $\alpha_l$ and $\alpha_t$. These parameters are considered known, their values are taken from the results of previous studies for the presented examples.

**Crosshole GPR**

The geophysical data used in this study are first-arrival travel times picked from crosshole GPR waveforms acquired between pairs of boreholes for which hydrological data were available. The radar travel times are dependent on the distribution of the radar velocities, a relation governed by the eikonal equation. The eikonal equation is again solved with the algorithm of Podvin and Lecomte (1991), that is, the same solver as for the hydraulic tomography. The travel times $t_{\text{GPR}}$ are described by a line integral along the ray trajectory:

$$t_{\text{GPR}} = \int_{\mathbf{x}_1}^{\mathbf{x}_2} u(s) ds, \tag{11}$$

where $u(s)$ [s/m] is the radar slowness (i.e., the reciprocal of the radar velocity) along the trajectory $s$, starting at point $\mathbf{x}_1$ and ending at $\mathbf{x}_2$.

# STRUCTURE-COUPLED JOINT INVERSION

## Data and models

We considered two application types of the joint inversion of geophysical and hydrological data. In the first application, GPR travel times were inverted jointly with both hydraulic pressure pulse travel times and hydraulic pressure pulse attenuation data. In the second appplication, GPR travel times were inverted jointly with tracer temporal moment data. In both cases, the data sets were inverted jointly with a structure-coupled approach, meaning that structural resemblance between the geophysical and the hydrological model (or models) is enforced by including a dissimilarity measure in the objective function. This approach is based on the formulation of Linde et al. (2006a), while we refer to Gallardo and Meju (2004) for an alternative formulation.

The forward problem consists of calculating the hydraulic and GPR travel times, the hydraulic attenuation and the temporal moments, from a given set of model parameters. In a very general sense, this is

$$\mathbf{d}_{\text{pred}} = \mathbf{F}(\mathbf{m}), \tag{12}$$

where $\mathbf{d}_{\text{pred}}$ contains the predicted data, calculated from the model parameter vector $\mathbf{m}$. For our applications, the forward operator $\mathbf{F}$ is derived from equations 2, 3, 6 and 11. In addition to the model parameters assigned to each pixel, we simultaneously invert for homogeneous background models of each model property. By doing so, convergence is more likely for cases when the initial model is poorly chosen.

For the first joint inversion application, the data vector is

$$\mathbf{d} = (\mathbf{t}_{GPR}; \mathbf{t}_{peak}; \mathbf{h}), \tag{13}$$

where $\mathbf{t}_{GPR}$ and $\mathbf{t}_{peak}$ are the GPR and hydraulic travel times and $\mathbf{h}$ the hydraulic attenuation data, and the model vector is

$$\mathbf{m} = (u_{ref}; \mathbf{u}; D_{ref}; \mathbf{D}; S_{s,ref}; \mathbf{S_s}), \tag{14}$$

where $\mathbf{u}$ is the radar slowness, $\mathbf{D}$ is the hydraulic diffusivity and $\mathbf{S_s}$ is the specific storage for all grid cells. Indices *ref* denote the parameters of the homogeneous background model. In the second joint inversion application, the data vector is

$$\mathbf{d} = (\mathbf{t}_{GPR}; \boldsymbol{\mu}_1), \tag{15}$$

where $\boldsymbol{\mu}_1$ are the normalized first temporal moments. The corresponding model vector is

$$\mathbf{m} = (u_{ref}; \mathbf{u}; \log(K_{ref}); \log(\mathbf{K})). \tag{16}$$

Note that *log* denotes the natural logarithm throughout this paper. The data combinations used for joint inversion are based on the availability of field data. Combining all three data types is straightforward in the inversion framework presented herein and recommended if the data are available.

**Sensitivity calculation**

The sensitivity matrices that relate the sensitivity of each data value to each model parameter are determined in different ways. For the temporal moment data, we calculate the sensitivities by the adjoint state method following Cirpka and Kitanidis (2000b). This method is based on determining performance functions (Sykes et al., 1985; Sun and Yeh, 1990) for all $N_\mu$ data, which are evaluated for all model parameters. For cases where $M \gg N_\mu$, as encountered in the inversion of the temporal moments, the adjoint state method requires much fewer forward calculations, compared to determining the Jacobian by individually calculating all partial derivatives of the data with respect to the model parameters, and is therefore more efficient (Cirpka and Kitanidis, 2000b). Concerning the GPR data, each travel time is linearly dependent on the slowness in the model cells that are trajected by the corresponding ray. The elements of the Jacobian are therefore given by calculating, for each datum, the ray lengths in each cell. The same applies to the sensitivities of the hydraulic tomography data, since we are treating the hydraulic tomography as a travel time problem.

**Data and model constraints**

The inverse problem aims at finding the most regularized models that predict the data within their error levels under the constraint of structural similarity. The objective function $\phi$ for $Q$ data sets is

$$\phi = \sum_{q=1}^{Q}(w_{d,q}\phi_{d,q} + w_{m,q}\varepsilon_q \phi_{m,q}) + \lambda \sum_{q=1}^{Q}\sum_{r<q}^{Q} T_{qr}. \tag{17}$$

Here, the factors $w_{d,q}$, $w_{m,q}$, $\varepsilon_q$ and $\lambda$ are weighting parameters, which will be described in detail later. The data misfit term $\phi_{d,q}$ for each data type $q$ is defined as

$$\phi_{d,q} = \left\| \mathbf{C}_{d,q}^{-0.5}(\mathbf{d}_q - \mathbf{F}_q(\mathbf{m}_q)) \right\|_p^p, \tag{18}$$

where $\mathbf{d}_q$ is the vector containing the measured data and $p$ is the order of the data residual norm (for the least-square misfit measure used herein, we have $p = 2$). $\mathbf{C}_{d,q}$ is the data covariance matrix, which under the assumption of unbiased, independent and Gaussian data errors, is diagonal and contains the data error variances (Menke, 1989), which include the measurement errors and possible errors in the forward calculation. In equation 17, $\phi_{\mathrm{m},q}$ is the model regularization term of the form

$$\phi_{\mathrm{m},q} = \left\| \mathbf{C}_{\mathrm{m},q}^{-0.5} (\mathbf{m}_q - \mathbf{m}_{q,\mathrm{ref}}) \right\|_2^2, \tag{19}$$

where $\mathbf{C}_{\mathrm{m},q}$ is the model covariance matrix and $\mathbf{m}_{q,\mathrm{ref}}$ is a reference model. For $\mathbf{C}_{\mathrm{m},q}^{-0.5}$, we use an anisotropic roughness operator that enforces the first-order differences between adjacent model parameters to be small. By explicitly penalizing roughness in the models, we target the smoothest model that still explains the data (e.g., Constable et al., 1987). To allow for sharper changes in the parameter field, we approximate an $L_1$-norm minimization for the model regularization term by applying iteratively reweighted least squares (IRLS) (e.g., Farquharson, 2008). For this purpose, $\mathbf{C}_{\mathrm{m},q}^{-0.5}$ is iteratively reweighted by multiplication with a diagonal matrix $\mathbf{R}_{\mathrm{m},q}$ with elements

$$R_{ii} = \left[ (m_{q,i} - m_{q,\mathrm{ref},i})^2 + \gamma^2 \right]^{-1/4}, \tag{20}$$

where the first term is taken from the previous iteration for every $i$-th grid cell and $\gamma$ is a small number used to ensure that $\mathbf{R}_{\mathrm{m},q}$ does not get singular for $\mathbf{m}_q \to \mathbf{m}_{q,\mathrm{ref}}$ $(\gamma = 10^{-4})$. The method applied can be seen as a modified $L_2$-norm which is why the $L_2$-norm formulation is kept in equation 19.

In equation 17, $T_{qr}$ is the sum of squares of the individual elements of the normalized cross-gradient function $\mathbf{t}_{qr}$ (Gallardo and Meju, 2004; Linde et al., 2008)

$$\mathbf{t}_{qr}(x,y,z) = \frac{\nabla \mathbf{m}_q(x,y,z) \times \nabla \mathbf{m}_r(x,y,z)}{\mathbf{m}_{q,\mathrm{ref}}(x,y,z) \cdot \mathbf{m}_{r,\mathrm{ref}}(x,y,z)}, \tag{21}$$

where $\mathbf{m}_q$ and $\mathbf{m}_r$ are two models of different model parameter types and $\mathbf{m}_{q,\mathrm{ref}}$ and $\mathbf{m}_{r,\mathrm{ref}}$ the corresponding reference models. In two dimensions, the enumerator in equation 21 simplifies to a scalar

$$t_{qr}(x,z) = \frac{\left(\frac{\partial m_q}{\partial z}\frac{\partial m_r}{\partial x} - \frac{\partial m_q}{\partial x}\frac{\partial m_r}{\partial z}\right)_{(x,z)}}{\mathbf{m}_{q,\text{ref}}(x,z) \cdot \mathbf{m}_{r,\text{ref}}(x,z)}. \tag{22}$$

Note that if more than two models are considered, equation 22 is applied to all possible model combinations (c.f., equation 17). We store the values $t_{qr}$ for every grid cell in a vector $\mathbf{t}$. To formulate a structural similarity constraint in the inversion, we first linearize the cross-gradient function at iteration $l$ (Gallardo and Meju, 2004)

$$\mathbf{t}^{l+1} \cong \mathbf{t}^l + \mathbf{B}^l(\Delta\mathbf{m}^{l+1} - \Delta\mathbf{m}^l), \tag{23}$$

where $\mathbf{B}$ is the Jacobian of the cross-gradients function with respect to the model parameters and $\Delta\mathbf{m}^l$ is the model update for the $l$-th iteration. The linearized cross-gradient function $\mathbf{t}^{l+1}$ is imposed to be close to zero. To do so, equation 23 is reformulated and included in the inverse problem so that models for which $\mathbf{t}^{l+1}$ is non-zero are penalized.

## The inverse problem

The non-linearity of the forward models and the cross-gradient function impose an iterative solution to the inverse problem. The resulting system of equations to solve at each iteration $l$ is

$$\begin{bmatrix} \mathbf{C}_d^{-0.5}\mathbf{J}^l & \mathbf{C}_d^{-0.5}\mathbf{J}_{\text{unit}}^l \\ \mathbf{C}_m^{-0.5} & 0 \\ \lambda\mathbf{B}^l & 0 \end{bmatrix} \begin{bmatrix} \Delta\mathbf{m}^{l+1} \\ \Delta\mathbf{m}_{\text{ref}}^{l+1} \end{bmatrix} = \begin{bmatrix} \mathbf{C}_d^{-0.5}\left[\mathbf{d} - \mathbf{F}(\mathbf{m}^l) + \mathbf{J}^l\Delta\mathbf{m}^l + \mathbf{J}_{\text{unit}}^l\Delta\mathbf{m}_{\text{ref}}^l\right] \\ 0 \\ \lambda(\mathbf{B}^l\Delta\mathbf{m}^l - \mathbf{t}^l) \end{bmatrix},$$

(24)

where $\mathbf{J}$ is the sensitivity matrix containing the sensitivities as described in the theory section on joint inversion and $\mathbf{J}_{\text{unit}}$ contains the sensitivities of the data to a constant change of the model throughout the model domain. For two data sets and models, we have the following data and model weighting matrices:

$$\mathbf{C}_d^{-0.5} = \begin{pmatrix} \varepsilon_q^l w_{d,q}^{0.5} \mathbf{C}_{d,q}^{-0.5} & 0 \\ 0 & \varepsilon_r^l w_{d,r}^{0.5} \mathbf{C}_{d,r}^{-0.5} \end{pmatrix}, \quad (25)$$

and

$$\mathbf{C}_m^{-0.5} = \begin{pmatrix} w_{m,q}^{0.5} \mathbf{R}_{m,q} \mathbf{C}_{m,q}^{-0.5} & 0 \\ 0 & w_{m,r}^{0.5} \mathbf{R}_{m,r} \mathbf{C}_{m,r}^{-0.5} \end{pmatrix}. \quad (26)$$

The model vector **m** includes all the individual models and is updated during each iteration step by adding the updates of the model $\Delta \mathbf{m}^{l+1}$ and the reference model $\Delta \mathbf{m}_{ref}^{l+1}$ to the original reference model $\mathbf{m}_{ref}$:

$$\mathbf{m}^{l+1} = \mathbf{m}_{ref} + \Delta \mathbf{m}_{ref}^{l+1} + \Delta \mathbf{m}^{l+1}. \quad (27)$$

Equation 24 is solved in a least-squares sense with the conjugate gradients algorithm *LSQR* (Paige and Saunders, 1982). The performance of the inversion process is quantified by the data fit for the individual data sets, which is formulated as a weighted root mean square error $RMS_q = \left( \frac{1}{N_q} \sum_{j=1}^{N_q} \frac{(d_j - d_{j,pred})^2}{\sigma_j^2} \right)^{0.5}$, with $\sigma_j$ denoting the standard deviation of the *j*-th datum.

## Data and model weighting

In the objective function (equation 17) and in equations 24-26, the factors $w_{d,q}$, $w_{m,q}$, $\varepsilon_q$ and $\lambda$ are weighting parameters. The weight given to an individual data set, $w_{d,q}$, is calculated based on the number of data points per data set to account for possibly large variations in the number of data between data sets through

$$w_{d,q} = \frac{1}{N_q}. \quad (28)$$

Since the data misfit term $\phi_{d,q}$ is expected to converge towards $N_q$ during the inversion process (i.e., a $RMS_q$ of 1 is approached), these data weights give similar weight to each data type. In the applications presented herein, weighting in this manner helped to adequately consider the hydrological data sets in the inverse modeling and to stabilize the convergence behavior during the inversion, which was challenging given that these data contained much less data points than the geophysical data sets. $w_{m,q}$ is the weight given to the regularization of the individual models. It is calculated from the model regularization matrix $\mathbf{C}_{m,q}^{-0.5}$ and the model update in the final iteration, $\Delta \mathbf{m}_q^{final}$ from previous individual inversions of the same data as follows

$$w_{m,q} = \frac{1}{\sum_{i=1}^{M}\left(\mathbf{C}_{m,q}^{-0.5} \cdot \Delta \mathbf{m}_q^{final}\right)^2} . \qquad (29)$$

This is the reciprocal of an estimation of the expected model misfit $\phi_{m,q}$ at the end of the inversion. By choosing the model weights like this, we account for the different magnitudes of variations in model properties expected in the different types of models. The trade-off parameter $\varepsilon_q$ defines the weight given to the regularization term with respect to the data misfit term. The optimal $\varepsilon_q$ is determined at each iteration step by a line search that seeks the $RMS_q$ closest to one. In the joint inversion, we sample from a plane (or from space in the case of three data sets) instead of a line to define the optimal combination of $\varepsilon_1$ and $\varepsilon_2$ (and $\varepsilon_3$). Determining $\varepsilon_q$ in this manner proved to be crucial in handling the vast differences in the amount of data and their information content, as well as the different convergence behaviors of the different model types. Finally, $\lambda$ represents the weighting of the cross-gradient function. It is typically assigned a rather large value ($10^3$-$10^5$) to force the cross-gradients function of the proposed models to be close to zero. Linde et al. (2008) showed that changes in λ around the optimal value have little influence on the resulting inverse models, it is therefore adequately represented by a uniform and constant value.

## SYNTHETIC EXAMPLE

The effect of the proposed joint inversion of hydrological and geophysical data on the estimation of the hydraulic conductivity was tested on a simplified synthetic aquifer profile. The artificial aquifer features two blocky hydrogeological facies embedded in a uniform background facies (Figure 1a), with the values of porosity, hydraulic conductivity and specific storage depicted in the figure.

The hydraulic diffusivities are determined using equation 1, while radar velocities $v_r$ for the three facies are derived from the porosity values by applying (in implicit form) (Sen et al., 1981)

$$\kappa = \kappa_w \varphi^m \left( \frac{1 - \frac{\kappa_s}{\kappa_w}}{1 - \frac{\kappa_s}{\kappa}} \right)^m , \qquad (30)$$

and (e.g., Davis and Annan, 1989)

$$v_r = \frac{c}{\sqrt{\kappa}} , \qquad (31)$$

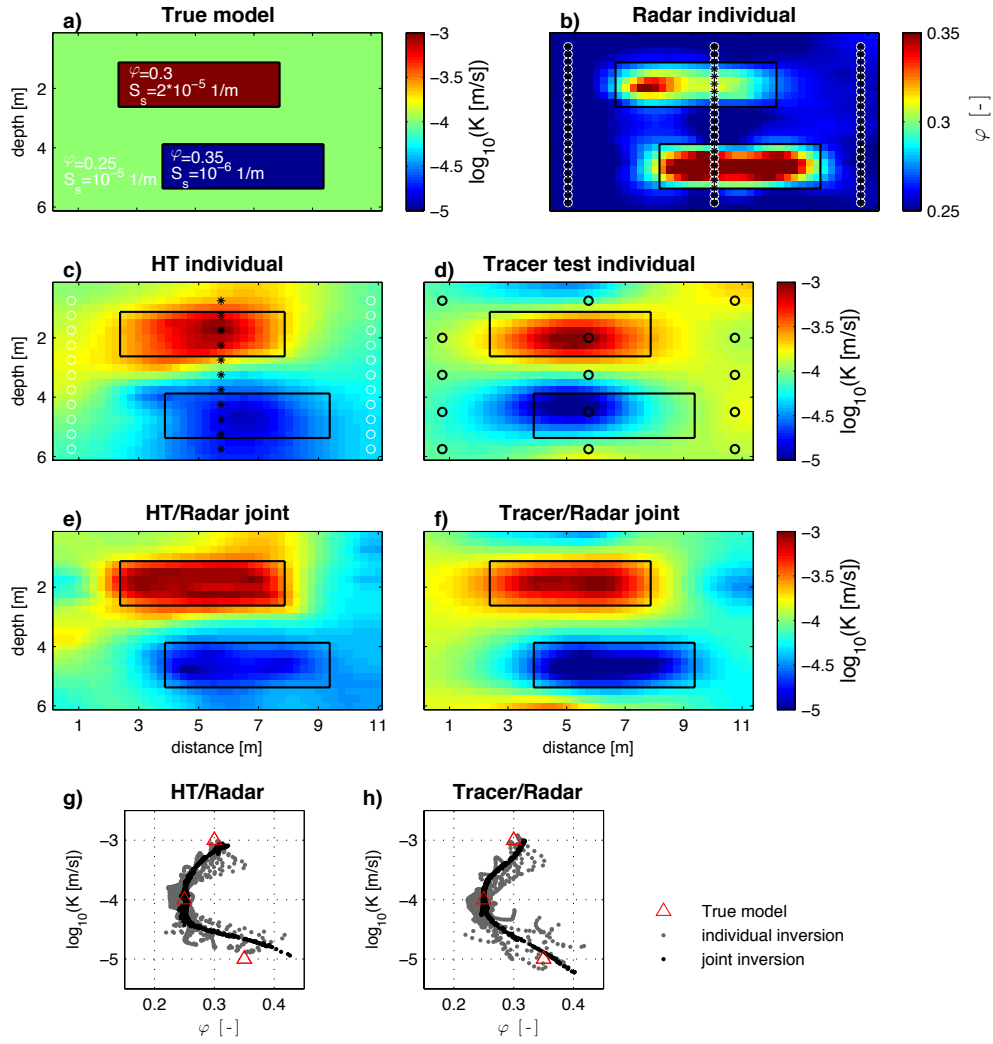

**Figure 1:** Inversion results for the synthetic example. a) True cross-section with porosity and specific storage values; color scale depicts hydraulic conductivity values. b) Porosity model, obtained by individual inversion of radar travel times. The velocity field is translated into porosities using equations 30 and 31. Transmitters and receivers are indicated by black asterisks and white circles, respectively. c) and d) Hydraulic conductivity models from individual inversions of hydraulic tomography data and tracer mean arrival times, respectively. Black circles denote tracer sensors. e) and f) Same as for c) and d) but for joint inversion with GPR travel times. g) and h) Scatter plots for the two different data combinations.

where $\kappa_w$, $\kappa_s$ are the relative electrical permittivity of water and the grains, respectively. $\kappa$ is the relative bulk electrical permittivity, $m$ is the cementation factor and $c$ is the speed of light. Synthetic data were calculated for all methods presented above, by solving the forward problems described through equations 2, 3, 6 and 11. The data were acquired in three boreholes, with a similar ray coverage and data density as in the field study described later. Note that for the hydraulic peak arrivals and attenuation data, as well as for the GPR travel times, we approximated the signals as incoming rays, instead of solving the diffusivity equation for the hydraulic tomography data and the Maxwell equations for the GPR signal. Since the purpose of this paper is to investigate the potential of structure-coupled joint inversions compared to individual inversions, we argue that including the full physics in the data generation is not necessary here. All data are contaminated with Gaussian noise of similar magnitude as the expected errors in the subsequent field study. The noise levels and all other relevant modeling parameters are listed in Table 1.

**Table 1:** Modeling parameters

| Parameter | Synthetic example | Field study |
|---|---|---|
| *Model geometry* | | |
| Discretization in *x/z*-direction | 0.25/0.25 | 0.25/0.25 m |
| Domain length in *x/z*-direction (profile A) | 11/6 m | 8/5.5 m |
| Domain length in *x/z*-direction (profile B) | | 30/5.5 m |
| *Parameters for tracer forward modeling* | | |
| Diffusion coefficient $D_m$ | $10^{-9}$ m²/s | $10^{-9}$ m²/s |
| Longitudinal dispersivity $\alpha_l$ | 0.1 m | 0.3 m |
| Vertical transversal dispersivity $\alpha_t$ | 0.001 m | 0.003 m |
| Mean head gradient $\Delta h$ | 0.002 m/m | 0.0033 m/m |
| *Parameters for porosity/velocity conversion* | | |
| Relative el. permittivity or water $\kappa_w$ | 81 | |
| Relative el. permittivity of grains $\kappa_s$ | 3 | |
| Cementation factor *m* | 1.5 | |
| Speed of light *c* | 3·10⁸ m/s | |
| *Initial models* | | |
| Radar slowness $u_{\mathrm{ref}}$ | 1/60 µs/m | 1/78 µs/m |
| Log hydraulic conductivity $K_{\mathrm{ref}}$ | -8 | -6 |
| Hydraulic diffusivity $D_{\mathrm{ref}}$ | 10 m²/s | 64 m²/s |
| Specific storage $S_{s,\mathrm{ref}}$ | 6.0·10⁻⁴ 1/m | 1.25·10⁻⁴ 1/m |
| *Weighting factor* | | |
| Cross-gradient weight $\lambda$ | 10³ | 10⁵ |
| *Error level* | | |
| Tracer mean arrival times | 5 % | 5 % + est. error* |
| GPR travel times | 1 % | 1 ns |
| Peak arrival time (hydr. tomography) | 5 % | 5 % |
| Attenuation (hydr. tomography) | 10 % | 10 % |

*This is the error estimate arising from the deconvolution to obtain virtual breakthrough data.

We first inverted the synthetic data sets individually following the procedure described in the theory section. Due to the dense spatial coverage of the radar experiment, the geometric shapes of the hydrogeological units are recovered best by the radar model (Figure 1b). The hydraulic tomography model (Figure 1c) does not image the facies blocks as sharply as the radar model, due to (1) higher error levels, (2) poorer ray coverage and (3) the fact that two tomograms (for hydraulic diffusivity and specific storage, not shown) are multiplied. The tracer arrival time inversion result (Figure 1d) fails to localize the facies blocks correctly, due to the limited data coverage. The hydrological data were then inverted jointly with the crosshole radar data, using data and model vectors as described in the theory section on inversion. The inverse models of hydraulic conductivity were improved in that the hydrogeological units are reproduced more accurately in terms of shape and location (Figure 1e and f). Scatter plots of hydraulic conductivity and porosity values recovered by the different inversions reveal the effect of the structural coupling in the joint inversions (Figure 1g and h). The estimated parameter values are concentrated along narrow lines between the true values for the joint inversions, whereas for the individual inversions, the estimated values are scattered around these lines. The estimated values between the true values are an effect of the applied smoothing regularization, for which gradual parameter variations between neighboring cells are an inherent property.

# INVERSION OF FIELD DATA

## Description of the Widen field site

The field experiments were conducted at the Widen field site in northern Switzerland. The site is located in the Thur River valley, with the closest borehole ~10 m from the river (Figure 2). Extensive alluvial gravel sediments of high hydraulic conductivity make the valley a major aquifer in the region (Jäckli, 2003). The aquifer body extends between depths of roughly 3 and 10 m and consists of sandy gravel (Naef and Frank, 2009). The aquifer is covered with a silty sand layer on the top and confined by lacustrine clayey silt at the bottom. The water table at the site usually is between 4 and 4.5 m depth and exhibits a hydraulic gradient of about 3.3 ‰, pointing away from the river flow direction at an angle of about 45° (Diem et al., 2010). An array of 18 boreholes aligned in a rectangular grid with distances of 3.5 m in north-south and east-west direction and 5 m diagonal distances has been set up (Figure 2a). All boreholes are cased over the aquifer thickness with fully-slotted PVC-tubes of either 5 or 11.4 cm diameter. The Widen site has been the object of several detailed geophysical and hydrological studies. The most relevant studies in the context of this paper are those of Doetsch et al. (2010), who presented results of 3-D joint inversions of GPR, seismic and ERT data and Coscia et al. (2011; 2012), who investigated the infiltration of river water into the aquifer by means of a 3-D ERT monitoring network.

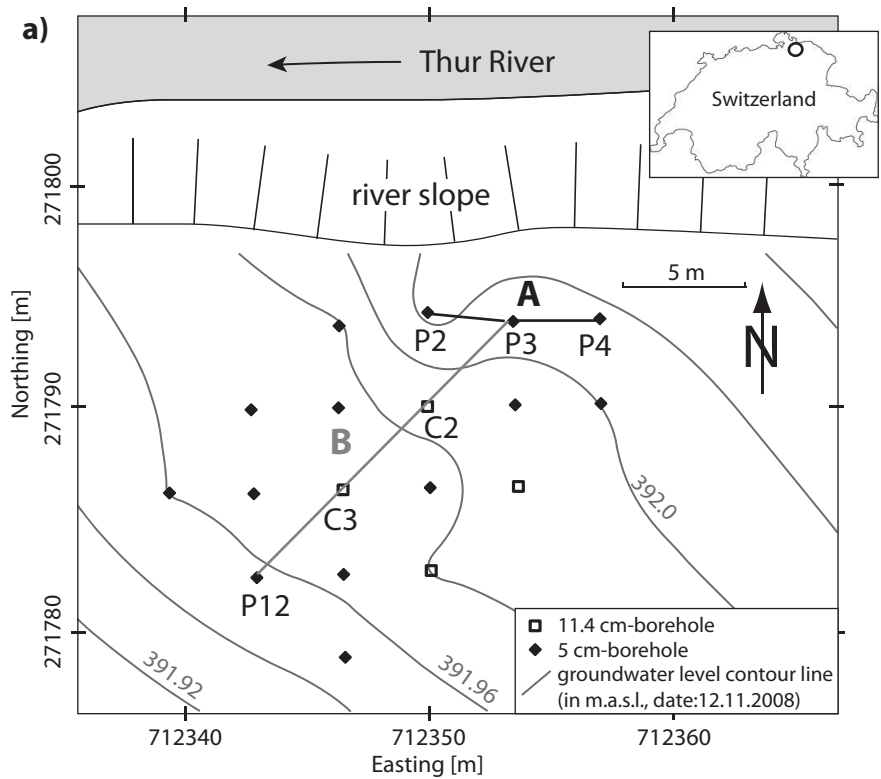

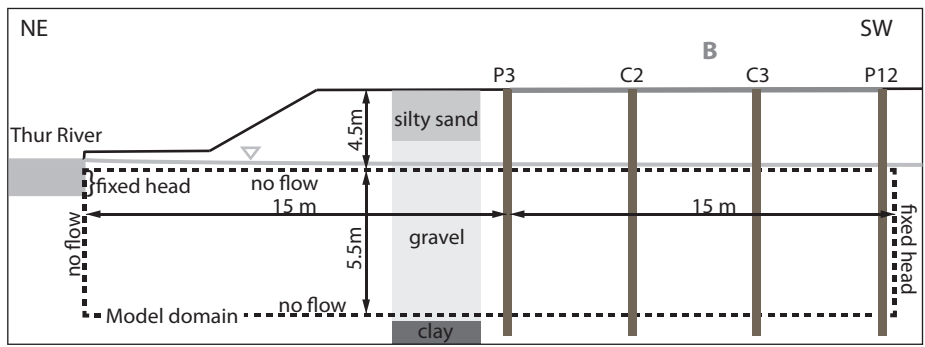

**Figure 2:** a) Overview of the Widen field site (modified after Diem et al. (2010)). Black squares and boxes depict 5 cm and 11.4 cm diameter boreholes, respectively. Gray lines indicate groundwater level contour lines. Coordinates are Swiss grid coordinates. We first applied our methodology to the planes below profile A (black). For this profile, we jointly inverted hydraulic tomography data and GPR travel times. Our second application was the joint inversion of tracer moments and GPR travel times, which were acquired in the planes below profile B (gray). b) Vertical cut along profile B, indicating the model domain (dashed box), the boundary conditions for the hydrological forward modeling and a schematic overview of the geological layers.

**Crosshole slug interference test data**

The crosshole interference slug test data were acquired between three wells (Figure 2a, profile A), where the center well was used as a test well and the two adjacent wells were used as observation wells. The tomographic set-up was realized by implementing double packer systems in the test and observation wells and varying their positions between 5.5 and 10 m, with one source gap at 8 m depth. The length of the injection and observation intervals was 0.25 m and the vertical spacing between them was 0.5 m. This experimental setup theoretically provides 180 travel time and attenuation data, of which 160 and 140 could be reliably used for travel time and attenuation analysis. The performance of crosshole interference slug tests in highly permeable unconsolidated sediments and its associated processing and hydraulic inversion is described in detail by Brauchler et al. (2007; 2010). The data could be fit to error levels of 5 and 10% on the hydraulic travel times and the attenuation data, respectively.

## Tracer mean arrival times

The tracer data were acquired along a profile parallel to the main groundwater flow direction (Figure 2a, profile B). Cirpka et al. (2007) developed a method to analyze bank filtration in a river-groundwater system by deconvolving time series of groundwater electrical conductivity measured in boreholes with those of the river water. They show that under the assumption of linearity and time-invariance, the transport between the river and a borehole can be described by a transfer function that relates the signal in the river *a(t)* to the signal recorded in the well *b(t)* by a convolution equation

$$b(t) = \int_0^T g(\tau) a(t-\tau) d\tau. \tag{32}$$

Here, $g(\tau)$ is the transfer function, $\tau$ is the travel time and *T* an upper boundary for $\tau$. The groundwater electrical conductivity signal therefore acts as a natural tracer. Since $g(\tau)$ describes the response to a pulse-like input, it can be interpreted as a breakthrough curve of a tracer injection in the river. At the Widen site, we had time series of electrical conductivity at three different depth levels available within three boreholes, recorded between May 2009 and March 2011, with sampling intervals of 15 minutes. These were processed analogous to time series published by Vogt et al. (2010), who applied the deconvolution method to data from a site nearby. We thus obtained virtual breakthrough data through deconvolution. For adjacent boreholes aligned parallel to the main groundwater flow direction, the mean arrival times of the transfer functions can be interpreted as the mean arrival time of a single pulse-injection tracer test in the river. In our analysis, we therefore considered the transfer functions from a borehole array aligned along the main direction of groundwater flow, which is estimated to be from northeast to southwest (Coscia et al., 2012). Extracting data from three boreholes and three depth levels ($z \approx 4.6, 6.6, 8.6$ m) resulted in nine mean arrival time data. To define a reasonable error level for these data, we considered the error estimates based on Cirpka et al. (2007). The determination of the transfer functions by deconvolution requires the minimization of a constrained objective function. The data and model errors can therefore be estimated by enforcing the data misfit to meet its expected value (Cirpka et al., 2007). To account for further uncertainties due to uncertain parameters in the above-mentioned optimization problem, we assumed an additional relative error of 5% on the data.

The forward calculations (equation 6) require estimates of the porosity, the dispersivities and the mean hydraulic gradient as additional input parameters. The porosity was calculated from the radar velocity field estimated by individual inversion of radar travel times using equations 30 and 31. The longitudinal and transversal dispersivities were taken from Doetsch et al. (2012), but the assumed dispersivities hardly affected the modeled mean arrival times. The detailed study of Diem et al. (2010) provided the assumed mean hydraulic gradient. All modeling parameters are listed in Table 1.

**Crosshole GPR data**

We conducted our first set of crosshole GPR measurements between three boreholes, along which hydraulic tomography data were available (Figure 2a, profile A). These boreholes are directed parallel to the river at a distance of ~3.5 m from each other. The second set of GPR data was acquired in the planes where tracer transfer functions were available. These planes are aligned along the main groundwater flow direction (Figure 2a, profile B), the boreholes are located at a lateral distance of 5 m. For all crosshole GPR experiments, we used a Malå 100 MHz slimhole transmitter/receiver system. Signals were recorded within the depth range of the gravel aquifer (~4 to ~10 m depth), with a transmitter and receiver spacing of 0.5 m and 0.1 m, respectively. Semi-reciprocal measurements were obtained by switching the transmitter and receiver boreholes (e.g., Doetsch et al., 2010). First-arrival travel times were first picked by hand before being refined automatically using a statistically-based information content picker (AIC picker, Leonard, 2000). For profile A, a total of 1493 travel times were obtained in two planes, while for profile B we obtained a total of 2283 travel times in three planes. The borehole deviations (Doetsch et al., 2010) were taken into account to accurately determine GPR transmitter and receiver positions, which was crucial to obtain reliable inversion results. The assumed error on the picked travel times is 1 ns, estimated from the summed errors caused by the time zero estimations (i.e., the time the electromagnetic signal needs to propagate through the electronics (Peterson, 2001)) and the picking error.

**Inversion of GPR and hydraulic tomography data**

In the first application of the joint inversion, we inverted GPR travel times and hydraulic tomography data. The model domain is a 2-D plane below profile A in Figure 2a between 4.5 and 10 m depth. We expect the 2-D assumption to be justified by the low subsurface heterogeneity revealed in previous studies (Coscia et al., 2011). The model discretization is 0.25 m in both *x*- and *z*-direction, which is in accordance with the vertical sampling resolution of the slug tests. The travel time-based forward calculations were performed on a grid that was refined by a factor of four in each direction. All inversions were regularized with an anisotropic roughness operator. We assumed an horizontal-to-vertical anisotropy ratio of 2:1, based on previous studies by Doetsch et al. (2010) and Diem et al. (2010). Different anisotropy ratios were tested, but produced similar results. The weight on the cross-gradient constraints $\lambda$ was $10^5$. For all models, homogeneous initial models (c.f., $\mathbf{m}_{q,\text{ref}}$ in the section on joint inversion) were chosen. The inversion parameters are summarized in Table 1.

For all three data sets (i.e., hydraulic travel times and attenuation data and radar travel times) the inverse models explain the data within the estimated error levels. The results of the individual and joint inversions are depicted in Figures 3a-f. All three types of model parameters (i.e., GPR velocity, diffusivity and specific storage) tend to be higher in the upper part and lower in the lower part of the aquifer. The hydraulic conductivity distribution is determined from the hydraulic diffusivity and the specific storage using equation 1. It is estimated to be around 0.005 m/s in the lower part of the aquifer and around 0.03 m/s in the upper part, with a smooth transition at ~7 m depth. In the individual inverse model of specific storage (Figure 3e), high and low values are concentrated in the center, which is probably an effect of the signal coverage decreasing with distance from the test well. This presumably unrealistic distribution is mapped into the hydraulic conductivity model (Figure 3g). In the conductivity model obtained by joint inversion, no such effect is observed and the transition between high and low conductivities is sharper (Figure 3h). Structural similarity is effectively increased between the joint inverse models. The value for the sum over the absolute values of $T_{qr}$ (c.f., section on joint inversion), which can be interpreted as a measure of structural difference, is decreased by a factor of ~1000, compared to the models obtained by individual inversions.

The relations between the GPR velocities and the hydraulic conductivities are investigated by a scatter plot of the corresponding values of each pixel (Figure 4a). Jointly inverting the different data sets leads to more focused relations between the models of GPR velocity and hydraulic conductivity, as enforced by the cross-gradient constraints. The models are, in general, positively correlated, except for a small region located in the upper left corner of the model domain, where the GPR velocity decreases with increasing hydraulic conductivity (see Figure 3b and h). This feature is recovered solely in the joint inverse models. As there is some radar ray coverage in this region and the individual inverse diffusivity model displays a significant increase in values, this feature is likely to be physically present and not an inversion artifact.

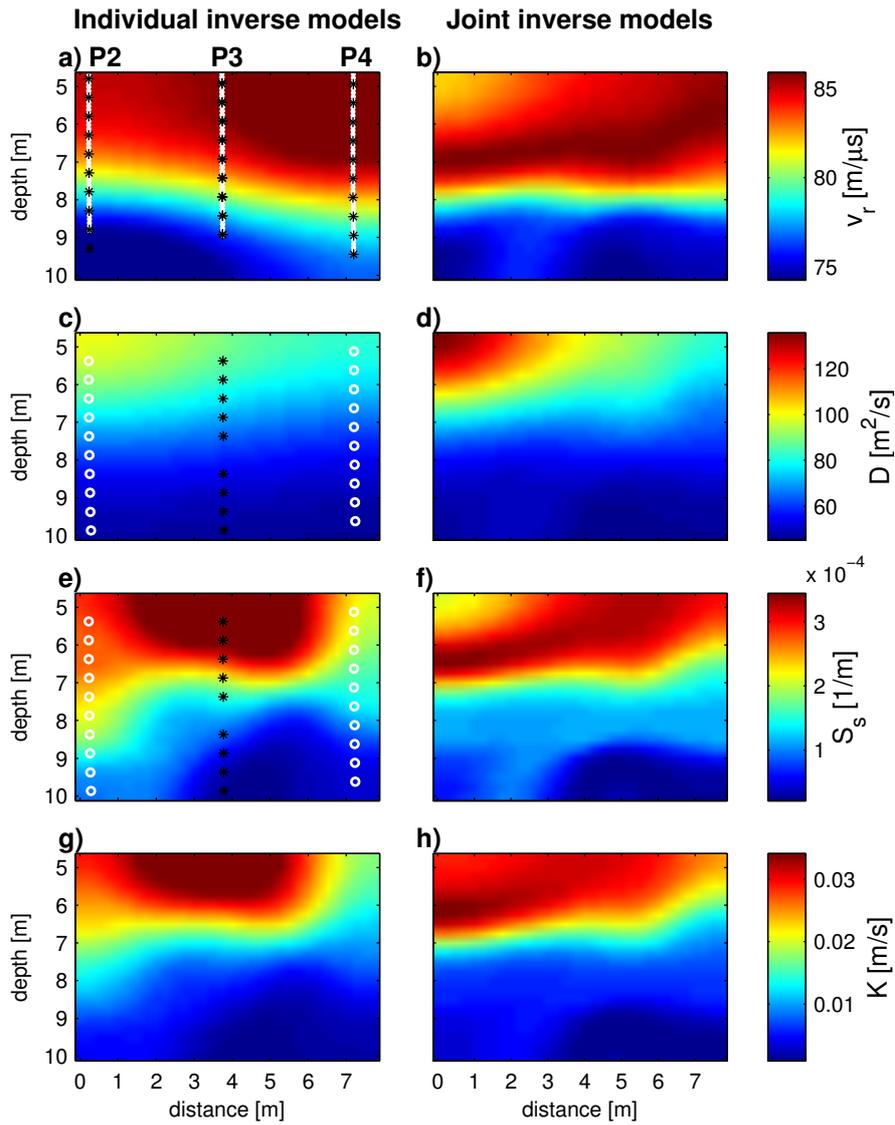

**Figure 3:** Results of the inversions of GPR travel times and hydraulic tomography data. Results from individual inversions are on the left, joint inversion results on the right. a) and b): GPR velocity models, c) and d): hydraulic diffusivity models, e) and f): specific storage models. The hydraulic conductivity models in g) and in h) are obtained by multiplying diffusivity and specific storage values for every grid cell. Black asterisks depict the positions of GPR and pressure pulse transmitters, white dots indicate the positions of GPR receivers and white circles the positions of the pressure sensors for hydraulic tomography. All models shown here predict the data with an RMS error of 1.02 or smaller.

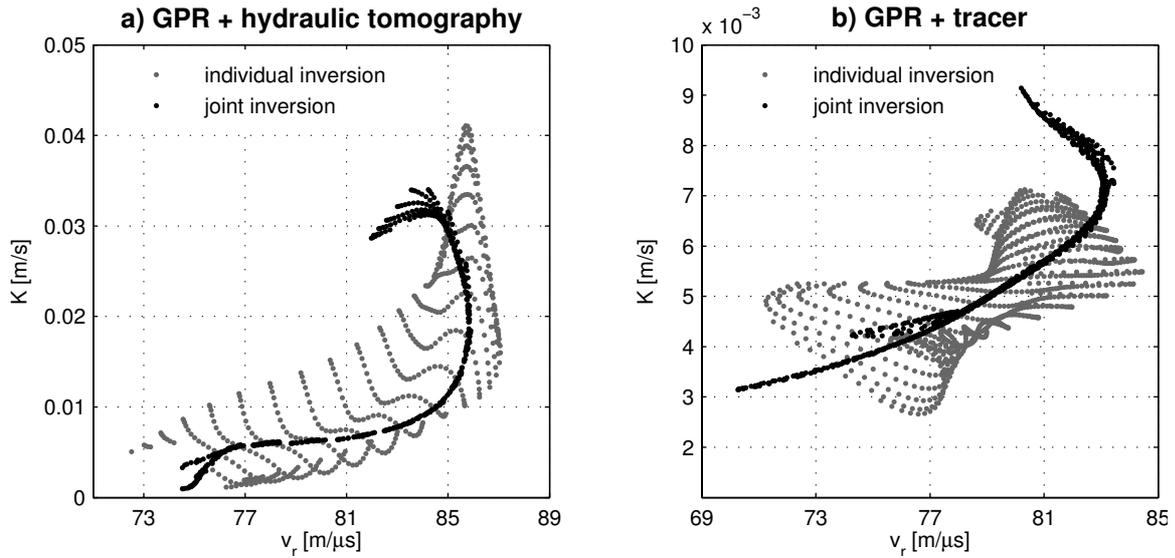

**Figure 4:** Scatter plots for the models obtained by a) inversion of GPR travel times and hydraulic tomography data and b) inversion of GPR travel times and tracer temporal moments. Gray and black dots indicate the models retrieved by individual and joint inversions, respectively. Model cells for which the sensitivities are zero are omitted.

## Inversion of GPR and tracer data

For the joint inversion of GPR and tracer data, the choice of the model dimensions and the boundary conditions for the hydrological forward modeling aims to adequately model the river water infiltrating the aquifer. Consequently, the model domain extends from the Thur River to the furthermost borehole in profile A in *x*-direction and from 4.5 to 10 m depth in *z*-direction (Figure 2b). The top and the bottom of the model are no-flow boundaries, as is the lower part of the riverside boundary. The upper 1.5 m of the riverside boundary, as well as the opposite boundary over the entire depth column, are fixed head boundaries to model in- and outflow of groundwater at these regions. The infiltration interval of 1.5 m is chosen to conceptually model river water infiltration into the aquifer. Varying the boundary locations and the infiltration interval over a realistic range had only minor effects on the resulting models in the borehole region. The distance between the boundary and the area of interest is therefore considered large enough to effectively decrease boundary effects in the region between sensors. Note that the area left of borehole C2 must be interpreted with care, since it is conditioned solely by the tracer arrival times measured in C2 and not by differences in arrival times as the rest of the domain. The head difference along the model domain is 0.15 m (i.e., a mean gradient of 3.3‰), as obtained from head measurements in the boreholes and extrapolated linearly to the riverside. We restricted our analysis to the part of the model below profile B, which is the area covered by the crosshole GPR experiments.

In the following, the same inversion parameters were chosen as for the joint inversion of radar and hydraulic tomography data described above. Individual inversions of the GPR travel times and tracer temporal moments provide a GPR velocity model (Figure 5a) showing the highest velocities at mid-depth around borehole C3. A rather sharp transition at about 7-8 m depth separates the upper part with higher velocities from lower velocities in the deeper part. The particularly low velocities in the deeper part between P3 and C3 are most likely related to a sand inclusion that has been localized around P3 (Diem et al., 2010; Coscia et al., 2011). The hydraulic conductivity model (Figure 5c) retrieved from the tracer data alone features conductivities decreasing vertically from around 0.006 m/s in the upper part to around 0.003 m/s in the lower part and very small lateral variations. When the GPR travel times and tracer data were inverted jointly, the GPR velocity model shows a more distinct low velocity zone in the upper part of the model domain between boreholes C3 and P12 (Figure 5b), compared to the individual inversion model. This region also shows high hydraulic conductivities in the model resulting from joint inversion (Figure 5d). In fact, this high conductivity/low velocity zone coincides with an interpreted zone of clean, well-sorted gravel, which is characterized by high porosities and flow velocities (Coscia et al., 2012; Klotzsche et al., 2012). This hydraulically conductive gravel zone, together with the low conductivity sand inclusion in the deeper part around P3, can explain the general increase of hydraulic conductivity with distance from the river and from the bottom of the aquifer to the top. The values lie in the range between 0.003 and 0.011 m/s. For this second application, the measure of structural differences between the models, $T_{qr}$, is decreased by a factor of ~600 for the joint inversion.

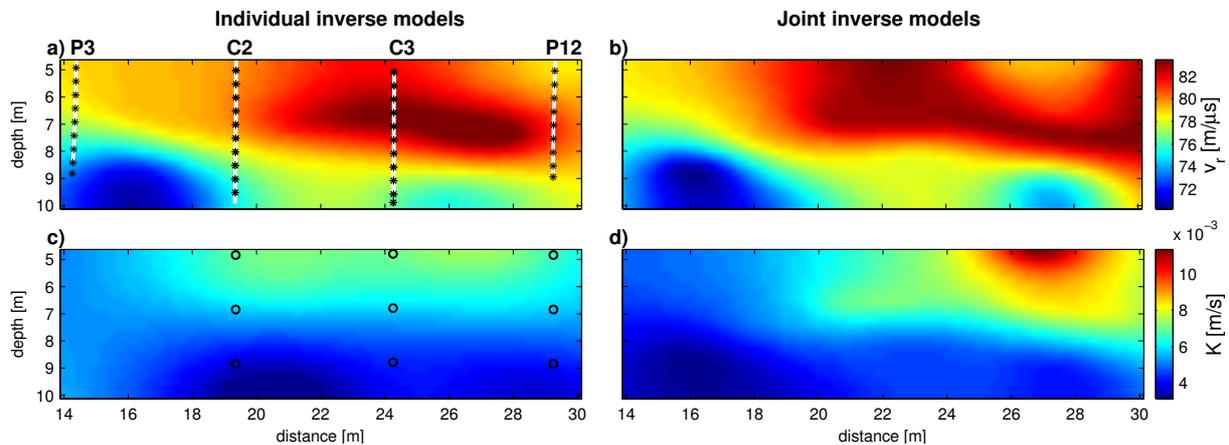

**Figure 5:** Inversion results of the inversions of GPR travel times and tracer temporal moments. Individual inversion results are shown on the left side, joint inversion results on the right. a) and b): GPR velocity models, c) and d): hydraulic conductivity models. Black asterisks and white dots in a) depict GPR transmitter and receiver positions, respectively. Black circles in c) depict the locations where the tracer signal was recorded. All models shown here predict the data with an RMS error of 1.03 or smaller.

The scatter between the models (Figure 4b) is similar to the previous application (c.f., Figure 4a). The GPR velocity and the hydraulic conductivity are in general positively correlated, but the joint inversion retrieves a region where this trend is reversed. This region is located in the upper right corner of the model domain, where the GPR velocity is decreasing with increasing hydraulic conductivity.

## Comparison of the inverse models to logging data

From the inversion results described above, we extracted the model parameters at the borehole locations P3 and C3. The GPR velocity models were converted into porosity values using equations 30 and 31 and compared to Neutron-Neutron probe data (2NUA-1000 probe, Mount Sopris Instrument Company, Inc.), which were calibrated to indicate porosities following Barrash and Clemo (2002). The individual and joint inversions provide very similar porosities at these locations (Figure 6a-b). Although overall variations are very small, the porosities obtained from the inversion models follow the general trend of the logging data.

At borehole C3, we compared the hydraulic conductivities obtained from the inversions to hydraulic conductivity values retrieved by flowmeter logging (Figure 6d). Flowmeter logs were acquired with an EM flowmeter device (9722 E-M Flowmeter, Century Geophysical Corp.). To get a depth profile of hydraulic conductivity, we measured both the ambient fluid flow in the borehole at intervals of 0.25 m and the flow while injecting water with rates of ~13 l/min. The difference of these flow measurements gave the net vertical flow rates, from which we calculated values of relative conductivity using its dependency on the vertical flow gradient (Molz et al., 1994; Paillet, 2000). The mean hydraulic conductivity has been estimated by a pumping test in borehole C2 to be 0.0148 m/s (Diem et al., 2010). Linear scaling of the relative values with this mean value provided the absolute conductivity values depicted by the dashed line in Figure 6d. Compared to the logging data, the inversion models underestimate the hydraulic conductivities by more than a factor of two. The variability in the inversion models is overly small, a result of the roughness regularization in the inversion and the small number of tracer data, but the joint inversion tends to increase the model variability. The general trend in conductivity versus depth is recovered in the inversion models, but not the small-scale variations.

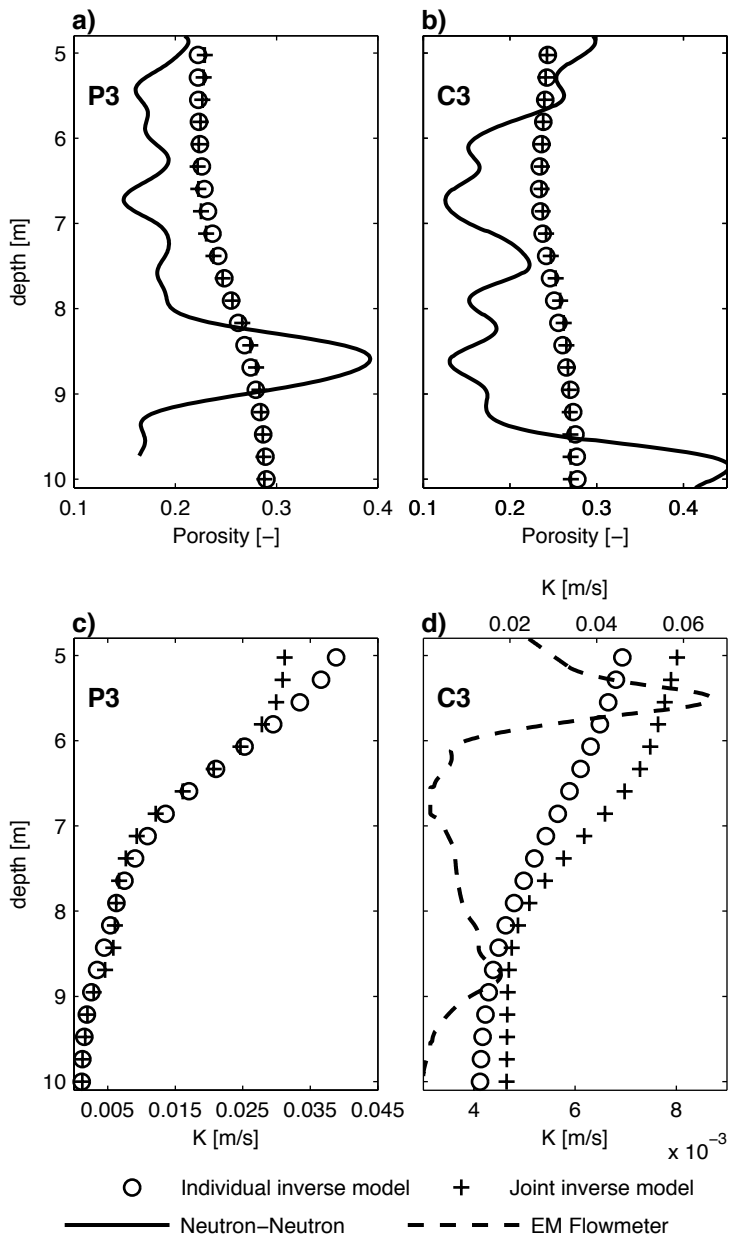

**Figure 6:** Inversion results in comparison with logging data. a) Porosity model at borehole P3 obtained by individual inversion (circles) and by joint inversion (crosses) of GPR travel times and porosity values from a Neutron-Neutron log (solid line). b) as for a), but for borehole C3. c) Hydraulic conductivity model at borehole P3 obtained by individual inversion (circles) and by joint inversion (crosses). d) as for c), but for borehole C3. The dashed line depicts hydraulic conductivity values obtained from EM flowmeter logging. Note the different scale for the flowmeter data (dashed line) on top of the box.

# DISCUSSION

Comparing the hydraulic conductivity models retrieved from the two applications, we encounter similar structures but differences in terms of absolute values. All models show lower conductivity values in the lower part and higher values in the upper part. The hydraulic conductivities retrieved by jointly inverting GPR and hydraulic tomography data (0.002-0.033 m/s) agree with the findings of previous studies. Diem et al. (2010) conducted a pumping test in borehole C2 and determined an effective hydraulic conductivity of 0.0148 m/s. Doetsch et al. (2012) monitored a salt tracer experiment for a nearby site using time-lapse 3-D ERT. Their estimate for the horizontal hydraulic conductivity is 0.04 m/s. Hydraulic conductivity estimates from joint inversion of GPR and tracer data do not exceed 0.011 m/s. Similar values of hydraulic conductivities are indicated by Coscia et al. (2012) from tracer velocity estimates and head measurements, as well as by multilevel slugtest results from borehole P3 (Diem et al., 2010). We have more confidence in the results from the joint inversion of GPR and hydraulic data and consider the conductivities obtained from joint inversion of GPR and tracer data an underestimation, since the tracer data are affected by the following issues: (1) The mean hydraulic gradient, which is an input parameter for our hydrological forward model, is extracted from interpolated piezometric measurements. This is a possible bias source, since head contours are not perfectly parallel and the gradients are small (c.f., Figure 2a); (2) To model the tracer arrival times as a function of the 2-D distribution of hydraulic conductivity between boreholes, the main groundwater flow direction must be parallel to the profile. We considered this a reasonable first assumption based on the head distribution (Diem et al., 2010) and estimates of flow velocity and direction by Coscia et al. (2012). However, heterogeneities in hydraulic conductivity can strongly influence the flow pattern, resulting in curved flow lines that are longer than the maximal distance in our model. For such flow lines we underestimate the hydraulic conductivity. Similarly, flow lines are longer than assumed (and hence conductivity is underestimated) if the true infiltration takes place further upstream and groundwater flow is not entirely parallel to the profile. (3) Another possible reason for the discrepancy in magnitude of the estimated conductivity fields for the two hydrological methods might be the effect of immobile domains (e.g., Haggerty and Gorelick, 1995; Singha et al., 2007). Small-scale heterogeneities with poorly connected pore space can result in retardation of the tracer, leading to long tails in the breakthrough curves. Since we are using the first normalized temporal moment of the breakthrough curves as tracer data, the measured mean arrival times are prone to being significantly higher if this effect is non-negligible.

The scatter plots of the results from the two joint inversion applications (Figure 4) do not indicate a unique relation between GPR velocity and hydraulic conductivity. Consequently, petrophysical coupling of the models during inversion might have led to biased results. At the Widen site, our results indicate that the two models are overall positively correlated. The model parameter relations retrieved by the two applications are very similar in shape, indicating that the assumption of structural similarity helped to recover the link between geophysical and hydrological properties. The differences between the individual and joint inversion results, that are the sharpening of the relation and a wider parameter range for the joint inversions, are analogous to previous applications using geophysical data only (e.g., Gallardo and Meju, 2003; Linde et al., 2008; Doetsch et al., 2010).

The results presented herein are adversely affected by the fact that the most permeable regions are found close to the water table, as also shown by flowmeter data and ERT monitoring results (Coscia et al., 2012). The GPR coverage is very poor in this region as a lot of data are discarded due to refractions at the water table. One solution to this problem that would also improve the resolution of the resulting velocity models would be to use full-waveform inversion. Recent algorithmic developments make it possible to include data acquired in the vadose zone, which allows to fully image the highly porous and permeable zone in the uppermost part of our models (Klotzsche et al., 2012).

The choice of model regularization strongly affects the inverse models. The combination of a smoothness regularization and sparse data coverage leads to smeared out features and potential loss of information regarding small-scale structures. Other regularization schemes, such as stochastic regularization (Maurer et al., 1998; Linde et al., 2006a), may better resolve small-scale structures but carry the risk of adding structures in model regions that are unconstrained by data where the prior model will be favored. Recently, the particular effect of model parameterization on the inversion of sparse tracer data has been analyzed in detail by Kowalsky et al. (2012). A possible way to circumvent the regularization effect could be to apply cross-gradients as a constraint in stochastic joint inversions.

The variations in hydraulic conductivity at the Widen field site are small. Assessing the joint inversion's potential to improve the estimated parameter fields is therefore difficult. The synthetic example in which property contrasts are higher, shows that structure-coupled joint inversions increase the ability to correctly infer the geometry of high- or low-conductivity zones, compared to individual inversions of hydrological data. It thus appears that benefits from applying the proposed methodology could be higher for systems that are more heterogeneous than in the present case study.

# CONCLUSIONS

The main objective of this study was to assess the applicability of structure-coupled joint inversions to combinations of geophysical and hydrological datasets. In a synthetic example we found that jointly inverting crosshole GPR data with either tracer arrival times or hydraulic tomography data under the assumption of structural similarity of the model parameters can improve the estimated hydraulic conductivity fields in terms of resolution and structure localization. Both types of joint inversion were applied to field data from a saturated gravel aquifer in northern Switzerland. With the two applications, we recover similar relations between the geophysical and hydrological model properties (i.e., GPR velocities and hydraulic conductivities). The model relations do not reveal a unique, easy-to-apply petrophysical link, which highlights the flexibility of the structure-coupled inversion approach compared with a petrophysical approach.

The estimated hydraulic conductivity values are in overall agreement with logging data and results from previous studies at the site. Hydraulic conductivity values retrieved by jointly inverting GPR and tracer data are lower than those obtained from joint inversion of GPR and hydraulic tomography data. This discrepancy may be related to uncertain input parameters and unaccounted 3-D effects in the forward model of the tracer mean arrival times. The recovery of small-scale variability as aspired by incorporating high-resolution geophysical data is limited by the choice of the model regularization, by the poor data coverage close to the water table where hydraulic conductivity is the highest, and by the sparse distribution of the hydrological data in particular. Nevertheless, the joint inversions provide hydraulic conductivity models that are more variable and in better agreement with auxiliary data than those obtained by individual inversions.

# ACKNOWLEDGEMENTS

This study was funded by the Swiss National Science Foundation (SNF) and is a contribution to the ENSEMBLE Project (grant no. CRSI22_132249). We thank Tobias Vogt for providing the tracer data, Wolfgang Nowak and Olaf Cirpka for their geostatistical inversion code and helpful comments, Tiffany Tchang for providing the logging data, Caroline Dorn for her help during GPR data acquisition and collaborators within the